\documentclass[10pt]{wlscirep}
\usepackage[utf8]{inputenc}
\usepackage[T1]{fontenc}
\usepackage{array}
\usepackage{booktabs}
\usepackage{longtable}
\newcolumntype{L}[1]{>{\raggedright\arraybackslash}p{#1}}
\usepackage{xcolor}
\usepackage{rotating}
\usepackage{hyperref}
\usepackage{textcomp}

\usepackage{todonotes}

\title{The Hidden Geometry of Global Aid: How Money Moves Through 10 Million Transactions}

\author[1,2,3,*]{Paul X.~McCarthy}
\author[1,2]{Xian Gong}
\author[2]{Marian-Andrei Rizoiu}
\author[4]{Paolo Boldi}
\affil[1]{League of Scholars, Sydney, Australia}
\affil[2]{University of Technology Sydney, Sydney, Australia}
\affil[3]{The University of New South Wales, Sydney, Australia}
\affil[4]{Università degli Studi di Milano, Milan, Italy}

\affil[*]{paul@leagueofscholars.com}

\begin{abstract}
International aid is usually described as flows between countries, hiding the organisations money passes through. We reconstruct global aid's organisational network from 10.9 million International Aid Transparency Initiative records, tracing over one million payments worth about US\$1 trillion between 22,773 organisations. First, centrality, measured as how far a funder’s money can travel downstream, is not size. Relative to their number of partners, giant bilateral donors (USAID, UK FCDO) are only as central as expected, while the most central positions belong to foundations (Gates, Hewlett), research organisations (ODI, University of Oxford) and delivery contractors (Mott MacDonald). Second, principal component analysis of eleven flow measures separates organisations along three lines: whether they mainly give or receive, how much money flows in, and how reliant they are on few partners. Humanitarian organisations emerge from the flows alone, tending to handle larger inflows. Third, almost all organisations form one connected network, yet money moves down long, one-way chains. Among five comparison networks, aid most resembles international banking, with a few large hubs feeding many small players, but has far fewer loops and is more strictly top-down. This suggests a system with little redundancy, where losing a major funder or intermediary could cut off organisations downstream, a pressing question after recent donor cuts.
\end{abstract}
\begin{document}

\flushbottom
\maketitle
%
%
\thispagestyle{empty}


\section*{Introduction}

International development finance has long been studied through questions of donor motivation, allocation, effectiveness, growth and poverty reduction\cite{dudley1976model,alesina2000gives,burnside2000aid,collier2002aid,djankov2008curse,easterly2006white}. It is commonly represented as a set of aggregate flows from donor countries or institutions to recipient countries. This representation is indispensable for measuring aid volumes, donor portfolios and geographic allocation, but it compresses the organisational infrastructure through which resources actually travel. In practice, a disbursement may pass from a government agency to a multilateral organisation, foundation, contractor, non-governmental organisation or local implementer before reaching its ultimate destination. The resulting system is therefore not only a collection of dyadic transfers but also a directed network in which position, reach and intermediary structure may matter independently of financial scale. This perspective connects aid research to work on donor fragmentation and coordination\cite{acharya2006proliferation,knack2007donor,torsvik2005foreign,aldasoro2010less}, and to a broader literature in which brokerage, weak ties and structural holes shape access to resources and opportunities\cite{granovetter1973strength,burt1992social}.

The international aid architecture has itself become more plural and organisationally complex, with bilateral agencies, multilaterals, philanthropic actors, non-governmental organisations and specialised intermediaries participating in overlapping governance arrangements\cite{fuhrer1994story,lancaster2008foreign,severino2009end,kharas2009new,stone2013knowledge,lewis2006development}. The Paris Declaration placed coordination, harmonisation and mutual accountability at the centre of aid-effectiveness reform\cite{oecd2005paris}. The International Aid Transparency Initiative (IATI) provides an unusual opportunity to study this infrastructure through a common transaction standard. IATI publishers report activities and transactions using common organisational identifiers and transaction types, making it possible, in principle, to reconstruct sender-receiver relations at organisational scale. Yet the data also contain a fundamental measurement problem: receiver identifiers are frequently absent. Missing counterparties are not random observations; they reflect publisher-specific reporting practices, and independent assessments have long documented that the completeness and quality of
IATI data vary widely between publishers \cite{ingram2014building, publishwhatyoufund2016index}. Treating unidentified receivers as ordinary missing data can therefore distort the actor universe, while merging them into a single anonymous node creates an artificial global hub. Any structural analysis must separate what is known about the wiring of identifiable organisations from what is known about the total volume of funding.

Previous studies have already shown that aid allocation and development assistance can be fruitfully analysed as networks\cite{coscia2013structure,swiss2016world,swiss2017foreign,horowitz2021rethinking,lim2008structure}. Network approaches have likewise illuminated global health aid, adaptation finance and humanitarian operations\cite{han2018social,fergus2022power,weiler2021donor,seybolt2009harmonizing,charles2022understanding}. Network science provides tools for making this distinction consequential, drawing on a mature literature on complex-network topology, small-world structure, robustness and network function\cite{watts1998collective,albert2000error,newman2003structure}. Shortest paths and harmonic centrality quantify downstream reach even in directed graphs with many unreachable pairs; centrality measures must be matched to the process or flow being studied\cite{boldi2014axioms,borgatti2005centrality}, while scalable neighbourhood-function methods provide related tools for very large graphs\cite{boldi2011hyperanf}. Degree-preserving randomisation can separate effects that follow mechanically from the number of partners from effects produced by higher-order wiring. Weighted and unweighted assortativity distinguish the topology of relationships from the concentration of money. Principal-component analysis (PCA) can identify latent organisational roles from flow features without introducing sector labels; complementary representation-learning approaches such as node2vec and UMAP offer alternative ways to recover network geometry\cite{grover2016node2vec,McInnes2018}. Finally, comparison with other directed flow networks can determine whether aid has a distinctive architecture or resembles familiar systems such as trade, banking, transport, migration or remittances.

Aid research has also developed strong causal and experimental traditions focused on whether particular interventions work\cite{banerjee2009experimental,duflo2006field,karlan2011more,glennerster2011small}. Our aim is different but complementary: rather than estimating the causal effect of a particular programme, we characterise the system-level architecture through which resources are routed. This system perspective is relevant to debates over aid coordination, global public goods, policy networks and the institutional reforms required for collective action\cite{ramalingam2013aid,lee2016transforming,moon2015will,shiffman2016emergence}.

Here we reconstruct the global IATI disbursement network and address three questions. First, which funders reach further through the network than their number of partners and their budget alone would predict (H1)? Second, what are the main patterns in how money moves between aid organisations, and do humanitarian and development organisations emerge as distinct groups from those patterns alone (H2)? Third, which other global flow systems, such as trade, banking or migration, does aid most resemble (H3)?

We call the property examined in H1 structural reach. Operationally, it is a funder’s outgoing harmonic centrality, the sum of inverse shortest-path distances to every organisation its payments can reach directly or through intermediaries, expressed relative to its expectation under a degree-preserving null model (see Methods). Structural reach is not brokerage in the sense of structural holes \cite{burt1992social} or betweenness centrality: it measures how far money can travel downstream from an organisation, not whether that organisation sits between otherwise disconnected parties.

To answer them we build three versions of the network from the same payment records. \textbf{IATI-K} (K for known recipients) includes only payments where both the sender and the recipient can be identified; we use it to trace how money travels from one organisation to the next. \textbf{IATI-ALL} adds the many payments whose recipient is not reported; we use it to measure how much each funder pays out in total. \textbf{IATI-CTY} groups payments by country, from the funder's home country to the country where the money is spent, so that aid can be compared with other country-level networks such as trade and banking. Separating the first two turns incomplete reporting from a hidden flaw in the data into something we can measure.

\section*{Results}

\subsection*{Aid money passes through long chains of intermediaries}
After restricting the  extract of data from IATI to disbursement transactions, removing exact duplicate rows and censoring a secondary reporter that republishes other organisations' activities, 3,643,369 candidate disbursement rows remained. Conservative identifier repair and receiver-name resolution yielded 1,330,963 rows for which both payer and receiver organisations could be identified. After canonicalising project-suffixed identifiers, removing self-loops and excluding placeholder identifiers, 1,051,065 transactions generated 28,708 distinct directed organisation-to-organisation relations among 22,773 organisations (Table~\ref{tab:data_construction}). Net disbursement value on these relations was approximately US\$1.03 trillion.

The graph is extremely sparse (density $5.54\times10^{-5}$) and highly asymmetric in organisational role. Only 852 organisations act as sources, whereas 22,378 appear as targets and 457 act as both. Nevertheless, 98.6\% of organisations belong to a single weakly connected component. Strong connectivity is radically different: the largest strongly connected component contains only 197 organisations (0.87\%). Thus, the aid system is not fragmented into isolated communities; rather, most organisations belong to one connected fabric whose edges overwhelmingly point in one direction.

\begin{figure}[ht]
\centering
\includegraphics[width=\linewidth]{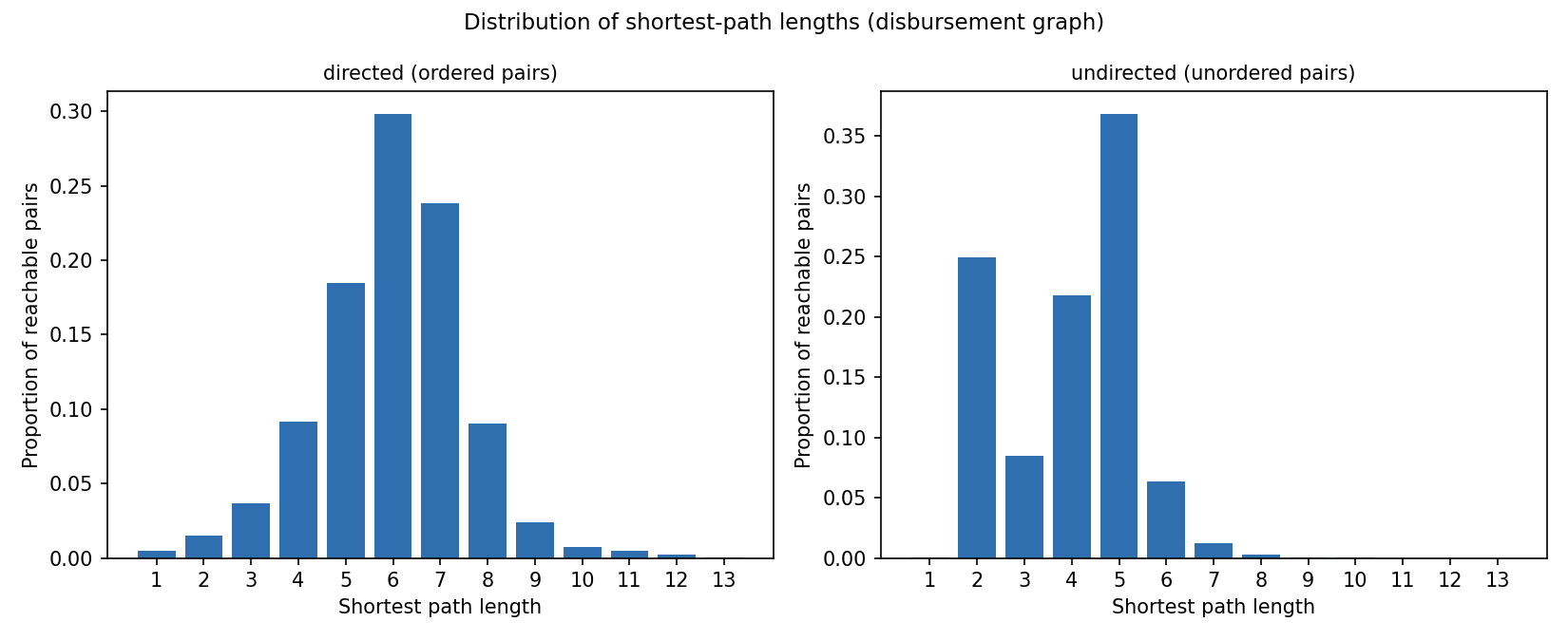}
\caption{\textbf{Money moves down long, one-way chains.} Following payments in the direction money flows (left), a route from one organisation to another usually takes five to seven steps, and such a route exists for only about 1\% of pairs. Ignoring direction (right), almost all organisations are linked, mostly within two to five steps, often because they share a funder or recipient. Shortest-path lengths (fewest payment steps between two organisations) in \textbf{IATI-K}. Left: pairs where one organisation can reach the other along the direction of payment. Right: the same network with direction ignored.}
\label{fig:distance}
\end{figure}

This hierarchy is also visible in the distance distribution. Only 1.11\% of all ordered pairs are connected by a directed path. Among reachable pairs, the mean shortest-path distance is 6.00, the effective diameter is 8 and the maximum observed distance is 13. More than 90\% of finite directed distances exceed two hops. When direction is ignored, by contrast, 97.28\% of unordered pairs are connected and the mean distance falls to 3.96. The symmetrised network has a large secondary peak at distance two, consistent with large sets of organisations sharing a common funder or receiver (Fig.~\ref{fig:distance}). These observations show that aid is simultaneously globally connected and strongly layered.

\begin{table}[h]
\centering
\setlength{\tabcolsep}{4pt}
\begin{tabular}{lr}
\toprule
Step & Records / value\\
\midrule
All records in IATI & 10,890,453\\
Disbursements (money actually paid out) & 3,741,876\\
After removing duplicate records & 3,661,084\\
Records republished by a secondary publisher, removed & 17,715\\
Recipient names matched to known organisations & 269,090\\
Payments with identifiable sender and recipient & 1,330,963\\
Payments to self (internal transfers), removed & 279,898\\
Payments used to build the network & 1,051,065\\
\midrule
\multicolumn{2}{l}{\emph{Resulting network (IATI-K)}}\\
Links between organisations & 28,708\\
Organisations & 22,773\\
Net value of payments & US\$1,030.4 billion\\
\bottomrule
\end{tabular}
\caption{\textbf{Constructing the organisational aid network from IATI records.} Starting from 10.9 million IATI transaction records, we retain disbursements, remove duplicates, republished records and internal transfers, and resolve sender and recipient organisations. The resulting IATI-K network contains 1,051,065 payments, forming 28,708 directed links among 22,773 organisations, with a net payment value of approximately US\$1.03 trillion.}
\label{tab:data_construction}
\end{table}

\subsection*{Reporting transparency changes the observable network}
Receiver identification is strongly publisher-dependent, in line with independent assessments of IATI reporting \cite{ingram2014building, publishwhatyoufund2016index}. We therefore constructed a companion graph, \textbf{IATI-ALL}, that retains every row with an identifiable provider. Disbursements lacking an identifiable receiver were routed to a provider-specific bookkeeping sink rather than to a single global ``unknown'' node. This preserves each provider's unreported downstream volume without asserting that anonymous receivers reported by different publishers are the same organisation. Because these nodes are pure sinks, they do not create additional paths among real organisations.

The broader graph contains 23,405 real organisations and 1,337 provider-specific unidentified sinks. Gross disbursement volume is US\$2.34 trillion, compared with US\$1.03 trillion in the identified network. The funder totals reported below (US\$2.61 trillion) are higher because they also count transfers within the same organisation, which are removed from the network. At the system level, 49.6\% of gross reported disbursement value has an identifiable receiver after name resolution. The median active funder identifies only 7.7\% of its disbursements, and 46.3\% identify no receiver at all. The large increase in transparency relative to an earlier extract demonstrates that a substantial share of apparent opacity resulted from receiver names being present without machine-readable identifiers.

The full funder-size distribution is extremely concentrated. Among 1,738 funders with positive disbursements, the median total disbursement is US\$0.37 million, whereas the ten largest funders account for 59.8\% of US\$2.61 trillion in gross disbursements and the top 1\% account for 76.0\%. A power-law tail fit gives $\alpha=1.37$ above US\$15.8 million, but a likelihood-ratio comparison does not reject a lognormal alternative ($R=-1.57$, $p=0.12$). Similar comparisons for degree in Table~\ref{tab:centrality_measure}, transaction strength and component sizes provide no consistent support for a pure power law, motivating degree-preserving rather than scale-free null models.

\subsection*{Cohesion depends on central funders, not the largest ones}
To test whether the network is held together by its largest funders or by its most central ones, we removed funders from IATI-K in decreasing order of gross US\$ disbursed, out-degree or outgoing harmonic centrality, and compared each ordering with removing the same number of funders at random (100 draws per k; Methods). Figure \ref{fig:precolation} tracks two measures: the share of organisations left in the largest weakly connected component, which ignores direction, and the share of ordered pairs still joined by a directed path, which follows money in the direction it flows.

Removing a single funder chosen by out-degree or by harmonic centrality (the same organisation heads both rankings) cuts the largest weak component from 98.6\% to 51\% of organisations and more than halves the number of directed reachable pairs. Removing the ten largest funders by US\$ value, among them the major bilateral agencies and development banks, still leaves 90\% of organisations weakly connected, only modestly below random removal of ten funders. About fifty top-US\$ funders must be removed before the weak component shrinks as much as it does after a single top-degree removal. The directed measure is more sensitive still. Removing the top three funders by out-degree leaves about 27\% of the original reachable pairs, by harmonic centrality about 43\%, and by US\$ value about 94\%. After seven removals all three rankings leave roughly 15\%, whereas random removal preserves most reachable pairs until several tens of funders are gone.

The network’s connectivity therefore rests on a few high fan-out funders and pooled channels rather than on the financial giants: cohesion follows degree and centrality, not money. Because the experiment deletes funders from the cumulative reported network and does not allow money to be rerouted, it measures structural dependence rather than predicting the effect of any actual funding cut.

\begin{figure}[ht]
\centering
\includegraphics[width=0.6\linewidth]{Figures/fig_percolation_v2.png}
\caption{\textbf{Removing central funders disconnects the network far faster than removing the largest.} Share of organisations remaining in the largest weakly connected component of IATI-K after removing the top k funders by gross US\$ (blue), out-degree (orange) or harmonic centrality (purple), or k random funders (grey; 100 draws per k). The leftmost point is the intact network; the horizontal axis is logarithmic.}
\label{fig:precolation}
\end{figure}

\subsection*{The most central funders are not the biggest}
To test H1, we measured each funder's outgoing harmonic centrality, $h(x)=\sum_y 1/d(x,y)$, on \textbf{IATI-K}. Harmonic centrality is appropriate because unreachable nodes contribute zero and no assumption of global strong connectivity is required. Financial scale was measured from \textbf{IATI-ALL}, so that a funder's total reported disbursement includes both identified and unidentified receivers. We then compared observed centrality with 960 degree-preserving random graphs generated by directed double-edge swaps. The resulting score
\begin{equation}
 z(x)=\frac{h(x)-\mu_{\mathrm{null}}(x)}{\sigma_{\mathrm{null}}(x)}
\end{equation}
quantifies multi-hop reach beyond that expected from the observed in- and out-degree sequence; this is what we term structural reach.

Because a funder with a single edge to a highly connected intermediary can inherit a large downstream neighbourhood, we restricted the headline ranking to organisations with out-degree at least five and gross disbursements of at least US\$1 million ($n=285$). Financial scale and harmonic centrality are positively associated (Spearman $\rho=0.65$; Kendall $\tau=0.47$), but scale explains only 29\% of the variance in centrality in an ordinary least-squares model using $\log_{10}$ disbursement value. More importantly, only 8 of the 50 organisations with the largest null-adjusted $z$-scores are also among the top 50 by financial scale.

\begin{figure}[ht]
\centering
\includegraphics[width=\linewidth]{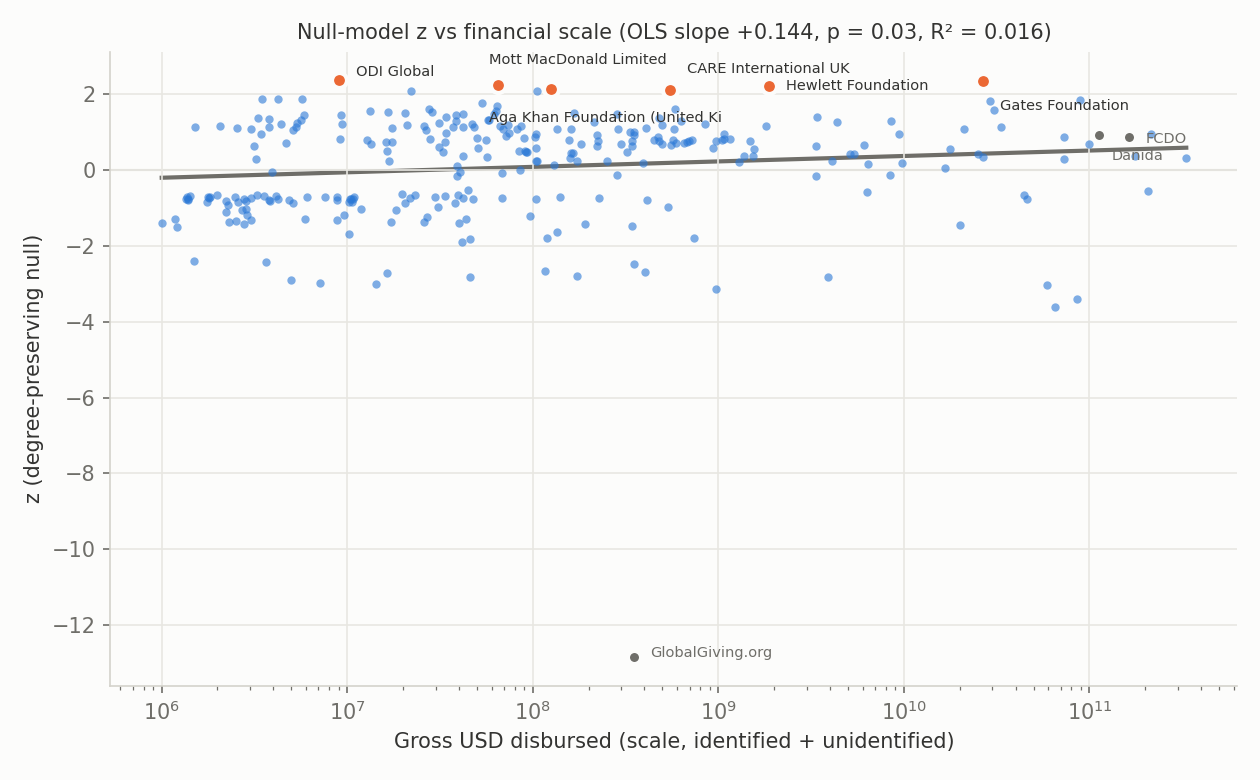}
\caption{\textbf{Bigger funders are not more central.} Each
dot is a funder, placed by how much it pays out (x-axis) and its adjusted
centrality (y-axis). The nearly flat line shows that money explains almost none of the difference: small and large funders spread across the same range.
GlobalGiving, far below the rest, pays many recipients directly. Funders with at least five recipients and US\$1m in payments ($n = 285$). Adjusted centrality is the $z$-score from Table~\ref{tab:h1top50}: raw centrality compared with randomised networks that keep each organisation's real number of partners. Payments include those to unidentified recipients; the horizontal axis is on a log scale.}
\label{fig:h1z}
\end{figure}

The upper tail consists mainly of foundations and specialist intermediaries (Table~\ref{tab:centrality_measure}). ODI and the Gates Foundation both have $z=2.37$, followed by Mott MacDonald ($z=2.26$), the Hewlett Foundation ($z=2.22$), Aga Khan Foundation UK ($z=2.15$) and CARE International UK ($z=2.12$). In contrast, GlobalGiving.org is an extreme negative outlier ($z=-12.8$): its raw harmonic centrality equals its out-degree because its many recipients lead to no additional downstream organisations. Its apparent reach is therefore almost entirely immediate rather than multi-hop. Large bilateral agencies are much closer to their degree-preserving expectation, including FCDO ($z=0.87$), Danida ($z=0.94$) and USAID ($z=0.31$). Once the degree sequence is controlled, regressing $z$ on financial scale yields $R^2=0.016$, indicating that less than 2\% of the residual variation is associated with scale (Fig.~\ref{fig:h1z}).

A looser null model that preserves only funder out-degree produces a similar ordering (Spearman $\rho=0.89$ between null-model scores) but much larger magnitudes. This inflation occurs because the looser model does not preserve popular receivers: it attributes to a funder both the advantage of attaching to highly connected intermediaries and any genuinely higher-order wiring. The degree-preserving swap null is therefore the more conservative test of structural surplus.

\begin{table}[p]
\centering
\scriptsize
\setlength{\tabcolsep}{4pt}
\begin{tabular}{rlrrrrrr}
\toprule
 & & \multicolumn{3}{c}{Size} & \multicolumn{3}{c}{Centrality}\\
\cmidrule(lr){3-5}\cmidrule(lr){6-8}
\# & Funder & Scale rank & Paid out (US\$m) & Recipients & Looser test ($z_{\mathrm{out}}$) & Raw ($h$) & Adjusted ($z$)\\
\midrule
1 & ODI Global & 218 & 9.0 & 5 & 60.61 & 4,435 & 2.37\\
2 & Gates Foundation & 19 & 26,785.7 & 226 & 8.60 & 7,969 & 2.37\\
3 & Mott MacDonald & 138 & 64.8 & 5 & 13.40 & 4,440 & 2.26\\
4 & Hewlett Foundation & 41 & 1,880.7 & 395 & 6.38 & 8,492 & 2.22\\
5 & Aga Khan Foundation UK & 112 & 126.0 & 6 & 51.85 & 4,441 & 2.15\\
6 & CARE International UK & 67 & 549.6 & 35 & 12.23 & 5,706 & 2.12\\
7 & London School of Hygiene \& Tropical Medicine & 187 & 22.1 & 7 & 118.02 & 4,440 & 2.10\\
8 & MSI Reproductive Choices & 115 & 105.5 & 23 & 19.35 & 5,824 & 2.09\\
9 & OCHA Pooled Fund (Chad) & 227 & 5.7 & 5 & 27.45 & 3,779 & 1.89\\
10 & ActionAid Rwanda & 247 & 3.5 & 5 & 27.67 & 3,654 & 1.88\\
11 & Leonard Cheshire Disability & 238 & 4.2 & 5 & 351.26 & 3,765 & 1.87\\
12 & US Dept of Defense & 8 & 89,283.5 & 11 & 76.65 & 4,766 & 1.84\\
13 & Gavi, the Vaccine Alliance & 18 & 29,248.0 & 59 & 10.95 & 5,742 & 1.83\\
14 & The Hunger Project Netherlands & 146 & 53.2 & 5 & 93.75 & 3,596 & 1.76\\
15 & US Dept of Commerce & 139 & 64.3 & 13 & 14.30 & 4,703 & 1.69\\
16 & Inter-American Foundation & 63 & 586.7 & 17 & 22.19 & 4,810 & 1.61\\
17 & International Development Law Organization & 181 & 27.7 & 5 & 76.37 & 3,254 & 1.61\\
18 & US Dept of the Treasury & 17 & 30,746.3 & 20 & 18.61 & 4,899 & 1.60\\
19 & University of Oxford & 140 & 63.0 & 6 & 56.85 & 3,726 & 1.57\\
20 & War Child Alliance Foundation & 204 & 13.3 & 8 & 15.05 & 3,978 & 1.56\\
21 & IUCN & 179 & 28.7 & 5 & 87.20 & 3,200 & 1.54\\
22 & Norwegian Refugee Council & 198 & 16.5 & 7 & 68.98 & 3,776 & 1.54\\
23 & SNV Netherlands Development Organisation & 102 & 166.1 & 7 & 34.73 & 3,799 & 1.52\\
24 & WYG International & 191 & 20.5 & 5 & 43.14 & 3,224 & 1.50\\
25 & World Vision UK & 90 & 285.0 & 14 & 15.75 & 4,463 & 1.49\\
26 & Woord en Daad & 141 & 61.5 & 5 & 103.98 & 3,250 & 1.49\\
27 & OCHA Pooled Fund (Niger) & 158 & 41.9 & 8 & 15.40 & 3,779 & 1.49\\
28 & Mensen met een Missie & 167 & 38.4 & 6 & 14.17 & 3,356 & 1.46\\
29 & International Fertilizer Development Center & 216 & 9.2 & 5 & 14.39 & 3,197 & 1.46\\
30 & United Nations Foundation & 226 & 5.9 & 5 & 21.77 & 3,294 & 1.46\\
31 & UK Dept of Health and Social Care & 38 & 3,412.5 & 108 & 8.55 & 5,728 & 1.40\\
32 & Cowater International & 171 & 33.9 & 8 & 61.36 & 3,726 & 1.40\\
33 & US Trade and Development Agency & 72 & 478.2 & 22 & 11.59 & 4,753 & 1.39\\
34 & SOAS University of London & 249 & 3.3 & 9 & 23.34 & 3,779 & 1.37\\
35 & University of Manchester & 244 & 3.8 & 8 & 16.39 & 3,694 & 1.36\\
36 & Terre des Hommes Netherlands & 142 & 58.0 & 7 & 21.21 & 3,412 & 1.34\\
37 & Wageningen University \& Research & 143 & 57.6 & 6 & 26.15 & 3,194 & 1.33\\
38 & ActionAid International Kenya & 228 & 5.6 & 6 & 120.28 & 3,216 & 1.32\\
39 & Adam Smith International & 166 & 38.6 & 7 & 51.86 & 3,293 & 1.30\\
40 & SOS Children's Villages Netherlands & 61 & 627.8 & 13 & 28.29 & 4,025 & 1.29\\
41 & UAE Ministry of Foreign Affairs & 28 & 8,577.4 & 10 & 25.04 & 3,814 & 1.29\\
42 & CARE Nederland & 97 & 215.3 & 11 & 19.42 & 3,862 & 1.28\\
43 & UK Dept for Business, Energy \& Industrial Strategy & 35 & 4,379.5 & 10 & 10.49 & 3,753 & 1.27\\
44 & WASTE & 229 & 5.4 & 5 & 34.44 & 2,941 & 1.24\\
45 & Frontline AIDS & 176 & 31.2 & 6 & 15.10 & 3,252 & 1.24\\
46 & Legacy Landscapes Fund & 151 & 47.3 & 5 & 77.98 & 2,764 & 1.23\\
47 & UTZ & 215 & 9.4 & 8 & 39.81 & 3,427 & 1.23\\
48 & University of Leeds & 236 & 4.4 & 10 & 53.67 & 3,679 & 1.23\\
49 & Disasters Emergency Committee & 55 & 843.2 & 28 & 11.94 & 4,492 & 1.21\\
50 & UK Dept for Energy Security \& Net Zero & 70 & 496.7 & 11 & 24.88 & 3,793 & 1.20\\
\midrule
\multicolumn{8}{l}{\emph{For comparison: the largest bilateral agencies and the lowest-scoring funder (position among all 285)}}\\
(156) & USAID & 1 & 332,452.2 & 302 & 4.66 & 5,252 & 0.31\\
(94) & UK FCDO & 5 & 162,694.4 & 1,143 & 3.62 & 8,006 & 0.87\\
(88) & Denmark MFA (Danida) & 6 & 112,467.7 & 568 & 4.18 & 6,986 & 0.94\\
(285) & GlobalGiving & 81 & 350.0 & 10,818 & $-8.25$ & 10,818 & $-12.84$\\
\bottomrule
\end{tabular}
\caption{\textbf{Top 50 funders by centrality in the global aid network.} Centrality here measures how many organisations a funder's money can flow on to, directly or through intermediaries, with nearer ones counting more. Each funder is scored against randomised networks in which every organisation keeps its real number of partners, so a high score means a more central position than its connections alone would give. Several small funders rank above the giant bilaterals. A central position shows how far money can travel. Funders with at least five recipients and US\$1m in payments ($n = 285$), ranked by $z$. USD is total payments, including to unidentified recipients. $z_{\mathrm{out}}$ uses a looser test that keeps only each funder's number of recipients, so its values run larger. Scale rank is rank by USD among the 285. Bottom rows show where the largest bilateral agencies and the lowest-scoring funder fall among the 285.}
\label{tab:h1top50}
\end{table}

\subsection*{Money movement primarily encodes organisational role and scale}
To test H2, we represented organisations using eleven value-aware structural features: log-transformed in- and out-degree, incoming and outgoing transaction counts, USD received and disbursed, mean incoming and outgoing transaction size, the fraction of transactions sent rather than received, and Herfindahl concentration indices for incoming and outgoing partners. PCA was performed on 4,711 organisations with at least ten transactions.

Table~\ref{tab:h2load} shows that the first four components explain 91\% of the variance. PC1 alone explains 50\% and contrasts outgoing with incoming activity, forming a clear funder--implementer axis. PC2 explains 24\% and primarily captures the scale of inflows. PC3 (11\%) and PC4 (7\%) are dominated by incoming and outgoing partner concentration, respectively. Thus, the principal dimensions of aid money movement are organisational role, scale and concentration rather than thematic sector.

\begin{figure}[ht]
\centering
\includegraphics[width=\linewidth]{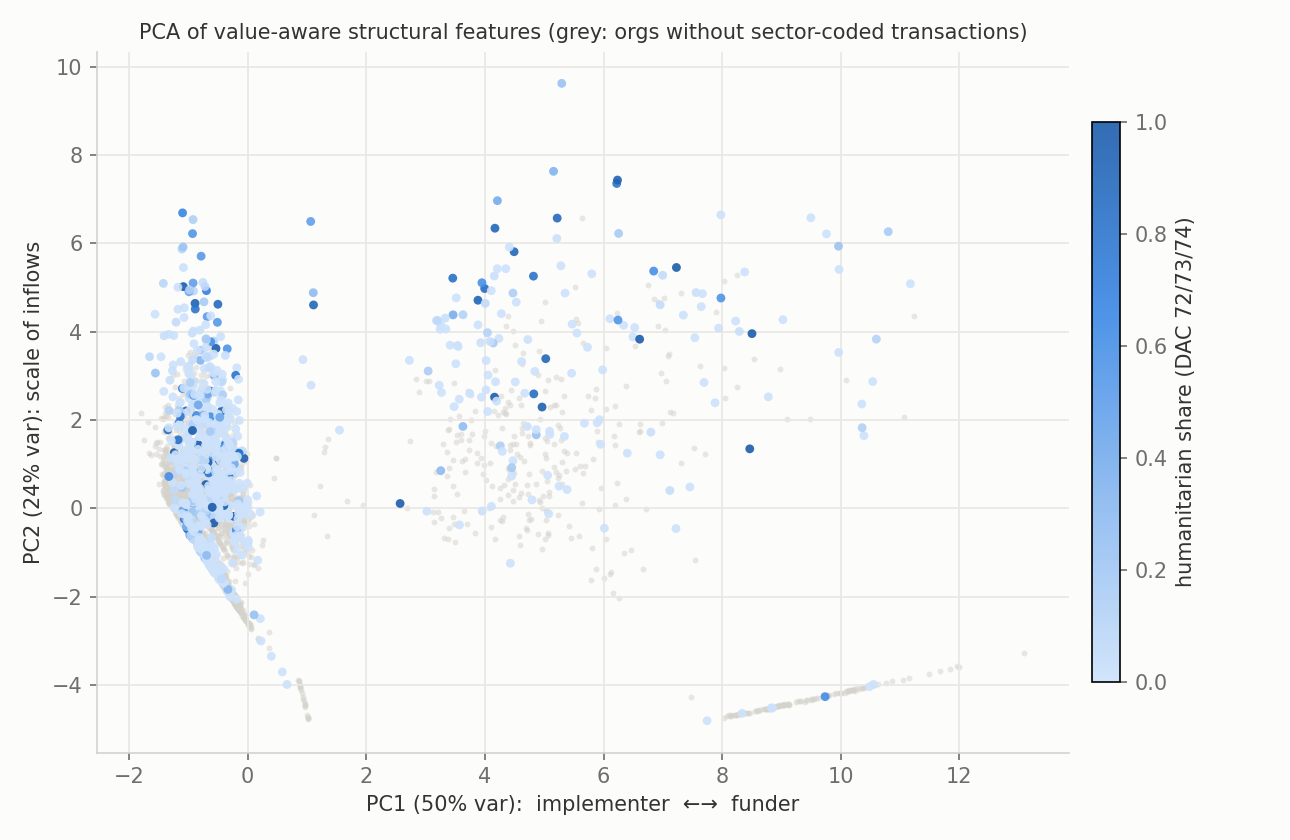}
\caption{\textbf{Organisations sort by role and size} Each dot is an organisation, placed using a statistical summary (principal component analysis) of eleven measures of how money moves through it. Across, the main pattern separates organisations that mainly give money (right) from those that mainly receive it (left). Up, the second pattern reflects how much money flows in. Colour shows how much of an organisation's work is humanitarian: darker dots sit slightly higher, but they mix with development organisations rather than forming a separate group. Organisations with at least ten transactions ($n = 4{,}711$); sector information was not used to place them. Humanitarian share is the proportion of sector-coded transactions in DAC codes 72--74 (emergency response, reconstruction and disaster preparedness); grey dots have no sector-coded transactions. The two axes capture 50\% and 24\% of the variation in the eleven measures. The thin line at bottom right is funders that receive no money themselves.}
\label{fig:h2}
\end{figure}

We externally validated this interpretation using the share of sector-coded transactions assigned to DAC humanitarian codes. Sector information was not used in the PCA. Among 788 organisations with at least five coded transactions, humanitarian share is moderately associated with PC2 (Spearman $\rho=0.28$, $p=9.3\times10^{-16}$) and only weakly with PC1 ($\rho=-0.11$). Humanitarian-dominant and development-dominant organisations separate on PC2 with AUC 0.67 (two-sided Mann--Whitney $p=8.0\times10^{-7}$). Humanitarian orientation is therefore detectable from flow structure, but it is a secondary signal rather than a principal axis (Fig.~\ref{fig:h2}).

\begin{table}[h]
\centering
\small
\begin{tabular}{lrrrr}
\toprule
 & Giving vs. & Money & Few or many & Few or many\\
 & receiving & coming in & funders & recipients\\
Measure & (PC1) & (PC2) & (PC3) & (PC4)\\
\midrule
Number of funders & 0.03 & \textbf{0.51} & \textbf{$-$0.44} & 0.04\\
Number of recipients & \textbf{0.38} & 0.12 & 0.15 & \textbf{$-$0.36}\\
Payments received & $-$0.09 & \textbf{0.50} & $-$0.12 & $-$0.01\\
Payments made & \textbf{0.40} & 0.11 & 0.15 & $-$0.24\\
Money received & $-$0.20 & \textbf{0.49} & 0.32 & $-$0.04\\
Money paid out & \textbf{0.41} & 0.11 & 0.15 & 0.02\\
Average payment received & $-$0.21 & \textbf{0.38} & \textbf{0.45} & $-$0.04\\
Average payment made & \textbf{0.40} & 0.10 & 0.15 & 0.14\\
Share of payments outgoing & \textbf{0.40} & $-$0.07 & 0.07 & $-$0.17\\
Concentration of funders & $-$0.23 & $-$0.20 & \textbf{0.61} & 0.00\\
Concentration of recipients & 0.25 & 0.07 & 0.14 & \textbf{0.87}\\
\midrule
Share of variation explained & 50\% & 24\% & 11\% & 7\%\\
\bottomrule
\end{tabular}
\caption{\textbf{What the main patterns in Fig.~\ref{fig:h2} are made of.} Each column is one pattern, and each number shows how strongly a measure drives it: near zero means not at all, larger values (positive or negative) mean more strongly, with the sign giving direction. The first pattern separates giving from receiving, the second reflects how much money comes in, and the last two capture whether an organisation relies on a few partners or spreads across many. Principal component loadings for the eleven structural measures; values of 0.35 or more in size are in bold. Together the four components explain 91\% of the variation.}
\label{tab:h2load}
\end{table}

\subsection*{Aid most resembles international banking, but is more top-down}
For H3 we compared aid with five publicly available directed weighted networks: international bank claims, merchandise trade, airline routes, refugee movements and bilateral remittances, summarised in Tables~\ref{tab:networks} and \ref{tab:datasets}. Because most comparison networks are country-level systems, the load-bearing comparison uses \textbf{IATI-CTY}, which aggregates IATI disbursements from the provider organisation's country to the activity's recipient country. Multilateral providers without a country attribution are excluded. The resulting network contains 224 countries and 2,028 directed arcs.

At matched scale, country-level aid is closest to the international banking network. Both have density 0.041, strong unweighted disassortativity (aid $-0.40$; banking $-0.44$) and almost identical directed reachability (12.0\% and 12.4\%, respectively). Trade is much denser (0.65) and highly reciprocal, while the airline network is almost symmetric. Refugee and remittance networks are substantially more reachable and cycle-rich.

Cycle-sensitive measures reveal an important difference even between aid and banking. Only 7.7\% of aid-country arcs participate in a directed cycle, compared with 29.7\% for banking. Directed two-paths close into three-cycles with probability 0.015 in aid versus 0.234 in banking. Trophic incoherence, which approaches zero for a perfectly layered hierarchy, is 0.043 for country-level aid and 0.033 at organisation level, compared with 0.290 for banking and much larger values for the other networks. Aid therefore resembles banking in sparsity, reachability and degree mixing, but is markedly more hierarchical.

Weighting links by money reveals a second layer of structure. Weighted reciprocity is smaller still (0.3\% of transaction weight is reciprocated at organisation level), but the sign of assortativity flips: unweighted mixing in IATI-K is $-$0.25, whereas weighted mixing is 0.25 by transaction count and 0.52 by US\$ value (Table \ref{tab:networks}). The flip is produced by a thin trunk of payments into the intermediary tier. The 2,319 organisation-level links whose recipient is itself a funder, about 8\% of all links, carry 28.9\% of identified US\$ value, mostly transfers between donor agencies and into large multilateral funds; excluding them lowers the US\$-weighted assortativity from 0.52 to 0.02. Below this trunk lies a large periphery: 91.0\% of recipients have exactly one identified funder. At country level, where multilateral intermediaries drop out by construction, weighted mixing is negative ($-$0.11), as expected of a redistributive donor-to-recipient layer.

Degree-preserving null models further distinguish organisation- and country-level structure in Extended Data Table \ref{tab:nullz}. Country-level aid lies relatively close to its own null for reciprocity, assortativity and average distance, suggesting that much of its global topology follows from which countries participate and how many partners they have (all \(|z|\leq6.3\)). International banking is statistically indistinguishable from its own null on all three measures (all \(|z|\leq1.5\)). Organisation-level aid, in contrast, is strongly non-random: reciprocity is 36 standard deviations above its already tiny null expectation and assortativity is 99 standard deviations more negative than expected. The distinctive wiring of aid therefore resides primarily in the organisational layer, where publisher-specific partner sets and intermediary pathways are visible.

\begin{table}[ht]
\centering
\small
\setlength{\tabcolsep}{4pt}
\begin{tabular}{lrrrrrrrrrr}
\toprule
Network & Nodes & Links & Density & Two-way & Two-way & Mixing & Mixing & Average & Reachable & Largest\\
 & & & & links (\%) & weight (\%) & & weight (\%) & steps & pairs (\%) & core (\%)\\
\midrule
Aid, organisations (IATI-K) & 22,773 & 28,708 & $5.5\times10^{-5}$ & 0.9 & 0.3 & $-$0.25 & 0.25 & 6.00 & 1.1 & 0.9\\
Aid, countries (IATI-CTY) & 224 & 2,028 & 0.041 & 1.2 & 0.02 & $-$0.40 & $-$0.11 & 2.35 & 12.0 & 11.6\\
\midrule
International banking & 170 & 1,169 & 0.041 & 25.5 & 50.2 & $-$0.44 & $-$0.02 & 1.68 & 12.4 & 12.4\\
Refugee movements & 203 & 4,745 & 0.115 & 10.8 & 6.2 & $-$0.45 & 0.51 & 2.96 & 80.4 & 79.8\\
Remittances & 213 & 10,227 & 0.227 & 46.8 & 14.3 & $-$0.31 & 0.13 & 1.81 & 81.8 & 81.7\\
Air routes & 3,425 & 37,594 & 0.0032 & 97.6 & 97.1 & $-$0.01 & 0.01 & 4.15 & 97.1 & 97.9\\
Trade & 186 & 22,451 & 0.652 & 88.5 & 67.8 & $-$0.32 & $-$0.08 & 1.35 & 100 & 100\\
\bottomrule
\end{tabular}
\caption{\textbf{Aid looks most like international banking, but its money flows one way.} We compare aid with five other global networks. At country level, aid and banking are equally sparse and equally hard to travel through, and in both, big hubs link mainly to small players. But very few aid links run both ways, whereas trade and air routes are almost entirely two-way. Density is the share of all possible links that exist. Two-way links is the share matched by a link in the reverse direction. Mixing (unweighted degree assortativity) below zero means well-connected nodes link mainly to poorly connected ones. Weighted columns use each network’s native weight (transaction counts for aid); two-way weight is weighted reciprocity \cite{squartini2013reciprocity} and weighted mixing correlates strengths rather than degrees. At organisation level the US\$-weighted mixing is 0.52. Average steps is over pairs that can reach each other. The largest core is the biggest group in which every member can reach every other, as a share of all nodes.}
\label{tab:networks}
\end{table}

\section*{Discussion}
The organisational network of global aid has a distinctive combination of global weak connectivity and strong directional hierarchy. Almost every organisation belongs to one weakly connected fabric, yet directed paths exist for only a small fraction of ordered pairs and the largest strongly connected core contains less than 1\% of organisations. This reconciles two intuitions that can otherwise appear contradictory: international aid is globally interconnected, but resources generally move down a layered chain rather than circulating through a densely reciprocal system.

This architecture makes structural reach different from financial scale. The distinction is consistent with classic network arguments that relational position can confer advantages not captured by actor size alone\cite{granovetter1973strength,burt1992social,borgatti2005centrality}, although our measure captures how far money can travel downstream rather than brokerage between otherwise unconnected actors. Large funders naturally have many immediate partners, and financial scale is positively correlated with harmonic centrality. However, the degree-preserving null model asks a stricter question: given exactly how many incoming and outgoing partners every organisation has, does a funder occupy a position that reaches unusually far downstream? Under this test, several foundations and specialist intermediaries outperform much larger bilateral agencies, whereas GlobalGiving provides the opposite example: enormous immediate breadth with almost no additional multi-hop reach. These results argue against using raw centrality as a proxy for influence. In a hierarchical flow network, centrality must be interpreted relative to degree and reporting structure. The removal experiment adds a complementary view. The null-adjusted score identifies funders whose reach exceeds what their number of partners implies, whereas the funders whose loss most disconnects the network are those with the largest raw fan-out; neither group coincides with the largest funders by value.

The organisations that rank highest fall into recognisable groups. Foundations such as Gates and Hewlett fund intermediaries that in turn fund many others. Research organisations such as ODI, the London School of Hygiene \& Tropical Medicine and the University of Oxford distribute money through research partnerships. Delivery contractors such as Mott MacDonald, Adam Smith International and Cowater manage programmes through chains of subcontracts. Pooled humanitarian funds, such as the OCHA country funds for Chad and Niger, and coordinators such as the Disasters Emergency Committee, gather money from many donors and pass it to responders on the ground. Some of the most striking cases are very small. ActionAid Rwanda paid out US\$3.5 million and Leonard Cheshire Disability US\$4.2 million, placing both in the bottom fifth of the 285 funders by size, yet they rank tenth and eleventh for centrality: their handful of recipients connect onward to a large part of the network. Such positions are consistent with filling a specialist need, whether geographic, as for a country office, or thematic, as for disability inclusion. Our data show where these organisations sit rather than how well they perform, but they suggest that small, specialised intermediaries may carry more of the system's connective weight than their budgets imply.

The latent-feature analysis reaches a complementary conclusion. Flow structure primarily distinguishes organisational role, scale and partner concentration. Humanitarian orientation is detectable but does not define a principal structural dimension. This is substantively important because it suggests that humanitarian and development organisations are not separate network species. Instead, they occupy partially different regions of a common organisational architecture, with humanitarian activity tending toward larger inflow scale. Because sector codes were withheld from the PCA, this association is an external validation rather than a feature-engineering artefact.

The cross-network comparison also refines how aid should be conceptualised. Comparative network reasoning is especially relevant in a world of interdependent systems, where topology can amplify fragility, systemic risk and abrupt transitions\cite{haldane2011systemic,vespignani2010fragility,helbing2013globally,scheffer2012anticipating}. At country level, international banking is the closest of the tested analogues in density, disassortativity and limited directed reachability. Both systems have restricted source sets and sparse cross-border relations. Yet aid is substantially more hierarchical and less cyclic, consistent with its redistributive function. Trade and air transport, by contrast, are intrinsically reciprocal systems and therefore poor structural analogues despite their importance as global networks. The resemblance between aid and banking should not be overstated: both datasets are reporter-censored, and part of their similarity may arise from observing a limited set of reporting source countries. The comparison is therefore structural rather than institutional.

The largest limitation is measurement. Counterparty identification depends on publisher practice, and the identified-receiver graph should not be interpreted as the complete aid system. Our two-graph design makes this limitation explicit: path-based conclusions apply to the identifiable organisational network, whereas volume and funder-scale conclusions use all reported disbursements with identifiable providers. Name-based receiver resolution substantially improves coverage, but conservative matching necessarily leaves some genuine organisations unresolved. Conversely, aggressive matching would risk creating false links. Future work could exploit receiver-reported incoming-fund transactions as an additional source of edges, probabilistic entity resolution, or longitudinal reporting changes to estimate the stability of the observed hierarchy.

A second limitation concerns financial weights. Currency conversion uses fixed indicative exchange rates rather than transaction-date rates, so USD values should be interpreted as approximate scale measures. Path analyses are deliberately unweighted and therefore unaffected by this choice. The degree-preserving null also conditions on the observed topology and cannot correct for edges that are entirely missing because of non-reporting. The network also pools disbursements reported between 1967 and 2025 into a single static graph, so a directed path may join payments made years apart. Path-based results describe the cumulative reported architecture rather than routes available at any one moment. Finally, the analysis is observational: structural reach describes potential downstream access in the reported disbursement network and should not be equated with causal influence, effectiveness or development impact.

Taken together, the results show that the global aid system is better understood as a directed organisational infrastructure than as a collection of bilateral totals alone. This interpretation complements complexity-oriented accounts of aid and international cooperation\cite{ramalingam2013aid,ostrom2017polycentric,sobel2025complexity,Tompson2023sustain} and research on transnational knowledge and philanthropic governance\cite{stone2010private,littoz2024knowledge}. Money concentrates in a thin trunk of payments among funding institutions, about 8\% of links carrying 29\% of identified value, above a large periphery in which 91\% of recipients have a single identified funder; organisational role and scale dominate its latent geometry; and some actors achieve disproportionate downstream reach without being the largest funders. Making this structure visible provides a basis for studying coordination, transparency, concentration and resilience in international development finance.

\section*{Methods}\label{sec:methods}

\subsection*{IATI data and transaction selection}
The analysis used the second transaction extract received on 8 September 2026, containing 10,890,453 rows. We retained IATI transaction type 3 (disbursements), yielding 3,741,876 rows, and removed exact duplicate rows. We excluded 17,715 rows published by AidData because inspection of activity identifiers and titles indicated that these records republished lending activities from other institutions. When the provider reference was absent, the reporting organisation was used as provider in accordance with the IATI transaction convention.

Organisation identifiers were normalised to uppercase. Identifiers consisting of an observed base organisation reference followed by a project suffix were collapsed to the base reference only when that base reference also occurred independently in the data. This existence guard avoids incorrectly truncating legitimate identifiers containing internal hyphens. Placeholder identifiers and self-loops were removed. A conservative repair pass corrected malformed country prefixes when the corrected identifier existed in the corpus and one-character provider variants when they were uniquely compatible with the reporting organisation and did not belong to dense numerical code families.

\begin{table}[ht]
\centering
\small
\begin{tabular}{l p{3.6cm} p{5.2cm} p{3.4cm}}
\toprule
Network & Source (snapshot) & A link from $x$ to $y$ means & Link strength\\
\midrule
Aid, organisations (IATI-K) & IATI payments (1967--2025) & organisation $x$ made at least one payment to organisation $y$ & number of payments (or US\$ value)\\
Aid, countries (IATI-CTY) & the same payments, grouped by country & an organisation based in country $x$ paid for an activity in country $y$ & number of payments (or US\$ value)\\
\midrule
International banking & BIS Consolidated Banking Statistics (2025 Q4) & banks based in $x$ hold claims on banks in $y$ & amount outstanding, US\$m\\
Trade & Correlates of War Trade v4.0 (2014) & $x$ exports goods to $y$ & value of goods, US\$m\\
Air routes & OpenFlights (2014) & at least one airline flies directly from airport $x$ to airport $y$ & number of airlines on the route\\
Refugee movements & UNHCR (end-2024) & refugees from $x$ are hosted in $y$ & number of refugees\\
Remittances & KNOMAD/World Bank (2021) & people living in $x$ send money home to $y$ & estimated flow, US\$m\\
\bottomrule
\end{tabular}
\caption{\textbf{The seven networks compared, and what a link means in each.} Each is a set of one-way links: a funder paying a recipient, banks lending across borders, goods exported, an airline route, refugees fleeing, or money sent home. Both aid networks come from the same IATI payments. The country version excludes multilateral funders, such as UN agencies and development banks, because they have no single home country. Remittances are model-based estimates rather than observed transfers. Full provenance is in Extended Data Table~\ref{tab:provenance}.}
\label{tab:datasets}
\end{table}

\subsection*{Receiver entity resolution and graph definitions}
Many disbursement rows contain a receiver name but no receiver reference. We constructed a receiver name--reference dictionary from two sources: IATI Registry publisher titles and within-corpus name--reference co-occurrences. A name was assigned to a reference only when at least three reference-bearing observations existed and one reference accounted for at least 90\% of those observations. Generic placeholders, redaction notices, role descriptions and other non-entity strings were blacklisted. This procedure resolved 269,090 previously reference-less receiver rows.

We analysed two related graphs. \textbf{IATI-K} contains only transactions for which both provider and receiver organisations are identifiable and is used for all path-based quantities. \textbf{IATI-ALL} contains every transaction with an identifiable provider; receiver-less disbursements are assigned to a provider-specific sink of the form \texttt{UNID::<provider>}. These sinks preserve unreported downstream volume while preventing unidentified receivers from different publishers from being merged. They are pure sinks and are excluded from interpretation as organisations. Gross funder scale and receiver transparency are computed from \textbf{IATI-ALL}.

The country-level network \textbf{IATI-CTY} aggregates the same disbursement records by provider country and activity recipient country. Provider country is derived from the organisation-reference country prefix or from the country associated with bilateral DAC agency codes. Multilateral providers without a unique country are excluded. The target is the transaction-level recipient-country code, falling back to the activity-level code when required.

\subsection*{Network measures}
For each graph we computed node and arc counts, density, in- and out-degree, transaction and value strengths, reciprocity, degree assortativity, weakly and strongly connected components, and shortest-path distributions. Unweighted reciprocity is the fraction of arcs whose reverse arc is present. Weighted reciprocity follows Squartini \emph{et al.}\cite{squartini2013reciprocity}, using $\sum_{ij}\min(w_{ij},w_{ji})/\sum_{ij}w_{ij}$ with negative net-value arcs clipped at zero. Directed assortativity is the Pearson correlation, over arcs, between source out-degree and target in-degree; weighted analogues replace degrees with strengths, and the US\$-weighted version is restricted to arcs with positive net value. In cross-network comparisons (Table \ref{tab:networks}), weighted measures use each network’s native weight: transaction counts for aid, US\$ millions for banking, trade and remittances, refugee numbers, and the number of operating airlines for air routes.

Exact shortest-path distributions were computed using breadth-first search over all sources. Harmonic centrality, a centrality measure with favourable axiomatic properties for disconnected networks\cite{boldi2014axioms}, for node $x$ is $h(x)=\sum_{y\neq x,d(x,y)<\infty}1/d(x,y)$. The effective diameter is the smallest distance containing at least 90\% of reachable pairs. Directed cyclic closure is defined as three times the number of directed 3-cycles divided by the number of open directed two-paths. Trophic incoherence $F_0$ was computed following MacKay \emph{et al.}\cite{mackay2020directed}, with values near zero indicating a layered hierarchy.

\subsection*{Degree-preserving null models}
For H1, null graphs preserve the complete directed in- and out-degree sequence. A proposed double-edge swap replaces $(a\to b,c\to d)$ with $(a\to d,c\to b)$ and is rejected if it creates a self-loop or duplicate edge. Twenty-four independent Markov chains were initialised from the observed graph. Each chain performed $10^{7}$ accepted swaps before the first retained sample and $2\times10^{6}$ accepted swaps between retained samples, producing 960 null graphs. For each funder, we calculated the mean and standard deviation of outgoing harmonic centrality across the null ensemble and converted the observed value to the $z$-score in equation (1). We also generated an out-degree-only null in which each funder's targets were sampled uniformly subject to a simple-graph constraint. This second null does not preserve receiver popularity and is used only as a robustness analysis.

The headline H1 analysis includes funders with observed out-degree $\geq5$ and gross disbursements $\geq$ US\$1 million. Associations between financial scale and centrality were quantified using Pearson correlation with $\log_{10}$ financial scale, Spearman rank correlation and Kendall's $\tau$. Ordinary least squares was used to estimate the variance in harmonic centrality explained by log financial scale. All reported tests are two-sided unless otherwise stated.

\subsection*{Funder-removal experiment}
Funders in IATI-K (the 852 organisations with at least one outgoing link) were ranked by gross US\$ disbursed (computed from IATI-ALL), by out-degree and by outgoing harmonic centrality on the intact graph. For each ranking and for k on a logarithmic grid from 1 to 852, the top k funders and all their links were deleted, and two quantities were computed on the remaining graph: the size of the largest weakly connected component as a share of the original number of organisations, and the number of ordered pairs joined by a directed path, as a share of all ordered pairs in the original graph, using exact breadth-first search. As a baseline, k funders were drawn uniformly at random, with 100 independent draws for each k. Per-k results are provided with the analysis code (percolation.tsv).

\subsection*{Latent structural features and humanitarian validation}
For H2, each organisation was represented by eleven structural variables. This unsupervised design follows the broader principle that structural features can reveal latent roles without imposing substantive labels in advance; network embeddings provide a related representation-learning perspective\cite{grover2016node2vec}. Specifically, each organisation was represented by: $\log_{10}(1+x)$ transformations of in-degree, out-degree, incoming transaction count, outgoing transaction count, USD received, USD disbursed, mean incoming transaction size and mean outgoing transaction size; the outgoing transaction ratio $t_{out}/(t_{in}+t_{out})$; and incoming and outgoing partner Herfindahl indices, $\sum_i p_i^2$. Features were standardised before PCA. Organisations with fewer than ten transactions were excluded, leaving $n=4,711$.

Humanitarian orientation was defined independently from DAC sector codes 72, 73 and 74. For each organisation, humanitarian share is the proportion of sector-coded transactions carrying one of these codes. External validation was restricted to organisations with at least five sector-coded transactions ($n=788$). Spearman correlation quantified association with continuous humanitarian share. For the binary contrast between humanitarian-dominant and development-dominant organisations, separation was summarised by area under the receiver-operating-characteristic curve and a two-sided Mann--Whitney test.

\subsection*{Comparison networks}
We processed five external directed networks using the same invariant and distance pipeline: BIS Consolidated Banking Statistics (2025 Q4, reporting-country banks to counterparty-country banks), Correlates of War bilateral trade (2014), OpenFlights scheduled routes (2014 snapshot), UNHCR refugee populations (end-2024), and KNOMAD/World Bank bilateral remittance estimates (2021). Edge weights represent outstanding bank claims, trade flow, number of airlines operating a route, refugee stock and estimated remittance flow, respectively. Because the systems differ in size and density, interpretation emphasises the country-level aid network and also reports deviations from each network's own directed degree-preserving null (96 double-edge-swap samples per network; Extended Data Table \ref{tab:nullz}). Null deviations are computed for unweighted invariants only, because arc weights have no canonical reassignment under swaps.

\subsection*{Software and reproducibility}
Graph construction and analysis were performed using custom Python and network-analysis code; exact all-pairs hop distributions were independently checked with a second breadth-first-search implementation. Power-law comparisons used the Clauset--Shalizi--Newman procedure as implemented in the \texttt{powerlaw} package\cite{clauset2009power}. Before submission, software versions, the public analysis repository DOI and a machine-readable construction log should be inserted here. The journal requires previously unreported custom code central to the conclusions to be available to editors and reviewers on request and encourages public archival of analysis code.

\section*{Data and Code availability}
The primary IATI transaction data come from the IATI Datastore and IATI Registry, and the comparison datasets are publicly available and summarised in Extended Data Table~\ref{tab:provenance}. All code used to construct the networks and reproduce the analyses, figures and tables is available at \url{https://github.com/boldip/AustralianAidNetwork}.

\section*{Acknowledgements}

We thank the International Aid Transparency Initiative (IATI) for maintaining and providing open access to the global aid data that made this research possible. We also acknowledge Snowmelt for its support and for valuable discussions that helped inform the broader context of this work.

\section*{Author contributions statement}

P.X.M. conceived the study, developed the core ideas, and led the manuscript. X.G. collected and curated the data and contributed to the empirical analysis and manuscript. P.B. designed and carried out the network analysis, including the null models, the cross-network comparison and the funder-removal experiment, and contributed to the interpretation. M.-A.R. provided conceptual guidance, methodological insights and critical feedback that shaped the analysis and interpretation. All authors reviewed and approved the final manuscript.

\section*{Additional Information}
The author(s) declare no competing interests.

\bibliography{sample.bib}

\newpage
\appendix
\renewcommand{\figurename}{Extended Data Fig.}
\renewcommand{\tablename}{Extended Data Table}
\setcounter{figure}{0}
\setcounter{table}{0}

\section*{Supplementary Information}


\begin{table}[ht]
\centering
\scriptsize
\setlength{\tabcolsep}{4pt}
\begin{tabular}{p{2.0cm}p{5.2cm}p{2.0cm}p{5.5cm}}
\toprule
dataset & source and URL & accessed / snapshot & construction notes and citation\\
\midrule
IATI transactions (\textsc{iati-k}, \textsc{iati-cty}, \textsc{iati-all}) &
\texttt{IATI\_All\_Transactions.par\-quet} retrieved from the IATI datastore API,
\url{https://api.iatistandard.org/datastore} & received 2026-09-08; 10{,}115{,}429 transaction rows & type-3 rows only; full-row dedup; AidData censored;
construction decisions in \hyperref[sec:methods]{Methods}; row counts at each step in Table ~\ref{tab:data_construction}\\[2pt]
publisher names & IATI Registry API,
\url{https://iatiregistry.org/api/3/action/organization_list} & 2026-09-02
& maps publisher refs to display names; used only to label funders in
Table~\ref{tab:h1top50}\\[2pt]
\textsc{ibank} & BIS Consolidated Banking Statistics, public bulk file
\url{https://data.bis.org/static/bulk/WS_CBS_PUB_csv_flat.zip} & 2026-09-02;
quarter 2025-Q4 & filter codes: measure S (stocks), bank type 4R (domestic
banks excl.\ domestic positions), basis U (ultimate risk), position C
(total claims), instruments A, maturity A, currency TO1, counterparty
sector B (banks); counterparties restricted to ISO-2 codes; USD millions.
Cf.\ Minoiu \& Reyes, J.~Fin.\ Stability 9(2), 2013\\[2pt]
\textsc{trade} & Correlates of War, Trade v4.0 (dyadic),
\url{https://correlatesofwar.org/wp-content/uploads/COW_Trade_4.0.zip},
file \texttt{Dyadic\_COW\_4.0.csv} & 2026-09-02; year 2014 (latest) &
arcs exporter$\to$importer, $w$ = import flow in millions of current USD
(\texttt{flow1}/\texttt{flow2}; $-9$ = missing and non-positive flows
dropped). Barbieri \& Keshk, COW Trade Codebook v4.0 (2016); Barbieri,
Keshk \& Pollins, CMPS 26(5), 2009\\[2pt]
\textsc{air} & OpenFlights route database, \texttt{routes.dat},
\url{https://github.com/jpatokal/openflights} & 2026-09-02; route snapshot
frozen June 2014 & nodes = airports; arc if $\ge 1$ airline operates the
route; rows with missing endpoint codes dropped\\[2pt]
\textsc{refugee} & UNHCR Refugee Data Finder population API,
\url{https://api.unhcr.org/population/v1/population} & 2026-09-03;
end-2024 stocks & arcs origin$\to$asylum country, $w$ = refugees (incl.\
refugee-like situations); unknown/stateless pseudo-origins dropped;
observational\\[2pt]
\textsc{remit} & KNOMAD / World Bank bilateral remittance matrix 2021,
\url{https://www.knomad.org} (retrieved via the Internet Archive; KNOMAD
discontinued 2024) & 2026-09-03; year 2021 & arcs sending$\to$receiving
country, $w$ = estimated flow in USD millions. Caveat: model-based
estimates from migrant stocks and incomes (Ratha--Shaw method), not
observed transfers\\[2pt]
\textsc{iati-cty} specifics & derived from the IATI parquet
(\texttt{07\_build\_country\_aid.py}) & as above & source country: ISO-2
prefix of the provider ref, or the country of a bilateral
\texttt{XM-DAC-<code>} agency (official OECD DAC donor codes; multilateral
codes have no country and are excluded); target country:
\texttt{transaction\_recipient\_country\_code}, falling back to the
activity-level code; the receiver-identified variant is kept as
\texttt{aid\_country\_k.tsv} for reference\\
\bottomrule
\end{tabular}
\caption{Provenance of every dataset used in this report: where each file
comes from, when it was retrieved, which snapshot is used, and how the
graph was constructed from it.}
\label{tab:provenance}
\end{table}

\begin{table}[h]
\centering
\setlength{\tabcolsep}{4pt}
\begin{tabular}{lrrll}
\toprule
What is measured & $\alpha$ & $x_{\min}$ & Test vs.\ lognormal & Verdict\\
\midrule
Funders per organisation (in-degree) & 2.46 & 2 & $R=-2.52$, $p=0.012$ & Lognormal fits better\\
Recipients per organisation (out-degree) & 1.57 & 1 & $R=-2.87$, $p=0.004$ & Lognormal fits better\\
Payments received (in-strength) & 1.81 & 27 & $R=-0.97$, $p=0.331$ & No clear winner\\
Payments made (out-strength) & 1.46 & 19 & $R=-1.65$, $p=0.099$ & No clear winner\\
Connected clusters, excl.\ giant (WCC) & 2.78 & 2 & $R=-0.50$, $p=0.620$ & No clear winner\\
Circulating groups (SCC) & \multicolumn{3}{l}{Almost all single organisations} & Too uniform to fit\\
\bottomrule
\end{tabular}
\caption{\textbf{Connections are highly unequal, but not in the textbook way.} Many networks are described as ``scale-free'', where a few hubs hold a vast share of all links. We tested this for aid and found no clear support: a lognormal pattern, with fewer extreme hubs, fits as well or better. So we judge each organisation against randomised versions of the real network, rather than against a standard hub-based model. Fits use the Clauset--Shalizi--Newman method (\texttt{powerlaw} package). The cluster-size fit excludes the one giant cluster (98.6\% of organisations) and rests on 93 small clusters; including it gives $\alpha = 2.50$, $x_{\min} = 2$. Sizes of circulating groups are too uniform to fit: almost every organisation is on its own, apart from a few small cycles and the 197-organisation core.}
\label{tab:centrality_measure}
\end{table}

\begin{sidewaystable}[p]
\centering
\small
\setlength{\tabcolsep}{6pt}
\scriptsize
\setlength{\tabcolsep}{9pt}
\begin{tabular}{lrrrrrrrrrrrrrrrr}
\toprule
network & $n$ & $m$ & \textbf{density} & \textbf{avg deg} & recipr. & recipr.$_w$ & \textbf{assort.} & assort.$_w$ & \textbf{avg dist} & \textbf{eff.\ diam} & reach. & \textbf{SCC} & \textbf{cyc.\ arcs} & $\boldsymbol{C_3}$ & $\boldsymbol{F_0}$ & $\Delta$\\
\midrule
\textsc{iati-k} & 22{,}773 & 28{,}708 & $5.5\times10^{-5}$ & 1.26 & 0.009 & 0.003 & $-0.25$ & $+0.25$ & 6.00 & 8 & 1.1\% & 0.9\% & 4.2\% & 0.011 & 0.033 & \\
\textbf{\textsc{iati-cty}} & 224 & 2{,}028 & $4.1\times10^{-2}$ & 9.05 & 0.012 & 0.000 & $-0.40$ & $-0.11$ & 2.35 & 4 & 12.0\% & 11.6\% & 7.7\% & 0.015 & 0.043 & \\
\midrule
\textsc{ibank} & 170 & 1{,}169 & $4.1\times10^{-2}$ & 6.88 & 0.255 & 0.502 & $-0.44$ & $-0.02$ & 1.68 & 2 & 12.4\% & 12.4\% & 29.7\% & 0.234 & 0.290 & 13\\
\textsc{refugee} & 203 & 4{,}745 & $1.2\times10^{-1}$ & 23.37 & 0.108 & 0.062 & $-0.45$ & $+0.51$ & 2.96 & 5 & 80.4\% & 79.8\% & 91.8\% & 0.064 & 0.288 & 15\\
\textsc{remit} & 213 & 10{,}227 & $2.3\times10^{-1}$ & 48.01 & 0.468 & 0.143 & $-0.31$ & $+0.13$ & 1.81 & 2 & 81.8\% & 81.7\% & 90.2\% & 0.375 & 0.622 & 23\\
\textsc{air} & 3{,}425 & 37{,}594 & $3.2\times10^{-3}$ & 10.98 & 0.976 & 0.971 & $-0.01$ & $+0.01$ & 4.15 & 6 & 97.1\% & 97.9\% & 99.9\% & 0.249 & 0.996 & 26\\
\textsc{trade} & 186 & 22{,}451 & $6.5\times10^{-1}$ & 120.70 & 0.885 & 0.678 & $-0.32$ & $-0.08$ & 1.35 & 2 & 100\% & 100\% & 100\% & 0.771 & 0.991 & 33\\
\bottomrule
\end{tabular}
\label{tab:network_measurement}
\caption{Same invariants, same code, seven directed networks. Four cycle/hierarchy columns: ``SCC'' is the largest strongly connected component's share of nodes; ``cyc.\ arcs'' the fraction of arcs whose endpoints lie in the same SCC, i.e.\ the arcs participating in some directed cycle; $C_3$ the cyclic closure, $3\cdot(\#\text{directed 3-cycles})/(\#\text{open directed 2-paths})$, the probability that a directed path $u\to v\to w$ is closed by the arc $w\to u$; $F_0$ the trophic incoherence of MacKay, Johnson \& Nicosia (R.~Soc.~Open Sci.\ 7:201138, 2020), 0 for a perfect layered hierarchy and near 1 for cycle-rich graphs. Rows are ordered by similarity to \textsc{iati-cty}: $\Delta$ is the sum, over the nine columns density, avg degree, assortativity, avg distance, effective diameter, SCC, cyc.\ arcs, $C_3$ and $F_0$, of the absolute difference between the row's rank and \textsc{iati-cty}'s rank in that column (average ranks on ties); \textsc{iati-k} and \textsc{iati-cty} are pinned first and shown without $\Delta$. ``reach.''\ is the fraction of ordered pairs at finite distance. Weighted columns use each network's native weight (transactions for aid, USD millions for interbank and trade, number of operating airlines for routes). Clauset LR tests against a lognormal reject or fail to favour a power law for the degree distributions of \emph{every} comparison network.}
\end{sidewaystable}

\begin{table}[h]
\centering
\large
\begin{tabular}{lrrrrrr}
\toprule
& \multicolumn{2}{c}{reciprocity} & \multicolumn{2}{c}{assortativity}
& \multicolumn{2}{c}{avg distance}\\
\cmidrule(lr){2-3}
\cmidrule(lr){4-5}
\cmidrule(lr){6-7}
network & value & $z$ & value & $z$ & value & $z$\\
\midrule
\textsc{iati-k} & 0.008 & $+36.1$ & $-0.253$ & $-98.5$ & 6.00 & $-0.3$\\
\textsc{iati-cty} & 0.012 & $+2.9$ & $-0.400$ & $+6.3$ & 2.35 & $+4.6$\\
\textsc{ibank} & 0.255 & $+1.1$ & $-0.440$ & $+1.5$ & 1.68 & $+0.3$\\
\textsc{trade} & 0.885 & $+36.0$ & $-0.315$ & $+11.2$ & 1.35 & $-0.3$\\
\textsc{air} & 0.976 & $+540.3$ & $-0.010$ & $+24.6$ & 4.15 & $+162.6$\\
\textsc{refugee} & 0.108 & $+10.0$ & $-0.453$ & $+4.0$ & 2.96 & $+9.5$\\
\textsc{remit} & 0.468 & $+31.4$ & $-0.310$ & $+24.4$ & 1.81 & $+16.2$\\
\bottomrule
\end{tabular}
\caption{Invariants as deviations from each network's own degree-preserving null. \textsc{ibank} is statistically indistinguishable from its null on all three invariants: given who reports and their degrees, its wiring is essentially random. \textsc{iati-cty} sits close to its null throughout (all $|z| \le 6.3$), with mild assortativity and distance deviations: who funds whom carries some real geographic--political pattern. \textsc{iati-k}, by contrast, carries strong genuine wiring (reciprocated arcs $36\sigma$ above chance despite the tiny raw value; disassortativity $99\sigma$ beyond what the degree sequence forces), and \textsc{air} is the most structured of all (route symmetry at $+540\sigma$; geographic communities stretching distances to $+163\sigma$).}
\label{tab:nullz}
\end{table}


\end{document}